\documentclass[prd,twocolumn,showpacs,reprint,preprintnumbers,nofootinbib]{revtex4-2}

\RequirePackage[colorlinks=true
,urlcolor=blue
,anchorcolor=blue
,citecolor=blue
,filecolor=blue
,linkcolor=blue
,menucolor=blue
,linktocpage=true
,pdfproducer=medialab
,pdfa=true
]{hyperref}

\usepackage{amsmath,amssymb,amsthm,amsfonts}
\usepackage{graphicx,tabularx}
\usepackage{color}
\usepackage{multirow}  
\usepackage{comment}
\usepackage{slashed}
\usepackage{enumitem}
\usepackage{cleveref}
\usepackage{rotating}
\usepackage[english]{babel}
\usepackage[justification=centerlast]{caption}
\usepackage{subcaption}
\usepackage{float}

\usepackage{breqn} 
\makeatletter 
\let\cref@old@eq@setnumber\eq@setnumber 
\def\eq@setnumber{%
\cref@old@eq@setnumber%
\cref@constructprefix{equation}{\cref@result}%
\protected@xdef\cref@currentlabel{%
[equation][\arabic{equation}][\cref@result]\p@equation\theequation}} 
\makeatother

\usepackage{cases}

\crefname{section}{Sec.}{Secs.}
\crefname{figure}{Fig.}{Figs.}
\crefname{equation}{Eq.}{Eqs.}
\crefname{appendix}{Appendix}{Appendices}
\setlist[description]{leftmargin=0.4cm}
\setlist[itemize]{leftmargin=0.4cm}

\newcommand{\be}{\begin{equation}\begin{aligned}}
\newcommand{\ee}{\end{aligned}\end{equation}}

\newcommand{\beq}{\begin{equation}}
\newcommand{\eeq}{\end{equation}}
\newcommand{\beqa}{\begin{eqnarray}}
\newcommand{\eeqa}{\end{eqnarray}}

\newcommand{\mev}{\text{MeV}}
\newcommand{\gev}{\text{GeV}}
\newcommand{\tev}{\text{TeV}}

\newcommand{\m}{\text{m}}

\renewcommand{\eqref}[1]{Eq.~(\ref{#1})}

\newcommand{\eg}{{\em e.g.}}

\newcommand{\TODO}[1]{\textcolor{green}{TODO}}
\RequirePackage[normalem]{ulem}

\DeclareUnicodeCharacter{2212}{\textendash}

\makeatletter
\def\l@subsubsection#1#2{}
\makeatother

\usepackage{multirow}
\usepackage{makecell}

\usepackage{fontawesome}

\begin{document}

\title{Looking forward to photon-coupled long-lived particles IV: neutralino-ALPino/gravitino}

\author{Krzysztof Jod\l{}owski}
\email{k.jodlowski@ibs.re.kr}
\affiliation{Particle Theory and Cosmology Group\char`,{} Center for Theoretical Physics of the Universe\char`,{} Institute for Basic Science (IBS)\char`,{} Daejeon\char`,{} 34126\char`,{} Korea}

\begin{abstract}
Various supersymmetric (SUSY) scenarios predict a sub-GeV neutralino decaying into a single photon and an invisible state. 
This signature has recently been studied in a number of intensity frontier experiments, finding constraints complementary to the usual collider searches.
In this work, we study the prospects of searches for long-lived neutralinos coupled to an ALPino or gravitino, where each can act as the lightest SUSY particle (LSP).
In addition to the neutralino decays into a LSP and a photon, we also consider three-body decays into a pair of charged leptons, and signatures related to scattering with electrons and secondary neutralino production.
For both models, we find that the searches at FASER2 will allow to overcome the current bounds, while SHIP will extend these limits by more than an order of magnitude in the value of the coupling constant. \href{https://github.com/krzysztofjjodlowski/Looking_forward_to_photon_coupled_LLPs}{\faGithub}
\end{abstract}

\renewcommand{\baselinestretch}{0.85}\normalsize
\maketitle
\renewcommand{\baselinestretch}{1.0}\normalsize

\section{\label{sec:intro}Introduction}

Searches for supersymmetric (SUSY) \cite{Golfand:1971iw,Gervais:1971ji,Ramond:1971gb,Neveu:1971rx,Wess:1974tw} particles with masses close to the electroweak scale has yielded no conclusive detection thus far \cite{ATLAS:2015vch,ATLAS:2017mjy,ATLAS:2022hbt,CMS:2015flg,CMS:2020bfa}.
This fact motivates the exploration of alternative possibilities, including the regime of low-mass neutralinos, possibly connected to other species by interactions beyond those predicted by the Minimal Supersymmetric Standard Model (MSSM) \cite{Dimopoulos:1981zb,Girardello:1981wz,Fayet:1974jb,Martin:1997ns}.

In particular, the neutralino of bino composition is almost unconstrained by collider searches because its couplings to gauge bosons vanish.
Moreover, if the neutralino decays, and therefore does not constitute dark matter (DM), its mass could be very light, possibly with masses in the sub-GeV range \cite{Gogoladze:2002xp,Dreiner:2009ic}.
Recent studies, \eg, \cite{Gorbunov:2015mba,deVries:2015mfw,Dercks:2018eua,Choi:2019pos,Dreiner:2022swd} have investigated such light neutralinos, which decay into SM states due to the R-parity violating interactions \cite{Allanach:2003eb,Barbier:2004ez,Dreiner:1997uz}.
Consequently, the bino behaves as a long-lived particle (LLP), making it especially well-suited for intensity frontier searches \cite{Battaglieri:2017aum,Beacham:2019nyx,Alimena:2019zri} looking for light and feebly-interacting particles beyond the Standard Model (BSM).

Another possibility is that the neutralino is in fact the next-to-lightest SUSY particle (NLSP), decaying into the LSP and a photon.
In this work, we study two such scenarios, both of them respecting the R-parity.

The first one corresponds to bino decaying into a SUSY partner of axion-like particle (ALP) called ALPino. This is an attractive BSM scenario that generalized the axion models invoked as a solution of strong CP problem \cite{Peccei:1977hh,Wilczek:1977pj,Weinberg:1977ma}, while the SUSY sector can solve the hierarchy problem \cite{Weinberg:1975gm,Gildener:1976ai,Veltman:1980mj,tHooft:1979rat} and provide a DM candidate \cite{Covi:1999ty,Covi:2001nw}.

The second scenario assumes that the LPS is composed of the gravitino, which is the fermionic partner of the spin-2 graviton. It is predicted by supergravity theories \cite{Volkov:1973jd,Fayet:1974jb,Deser:1977uq,Freedman:1976xh}, which incorporate local supersymmetry transformations.
For consistency, they necessarily combine SUSY and general relativity, while at the same time mitigating the hierarchy problem. Moreover, in certain regime, gravitino can be a DM candidate \cite{Ellis:1983ew,Ellis:1984eq,Steffen:2006hw}.

We note that the displaced bino decay signature has been investigated in multiple previous works.
In particular, ref. \cite{Choi:2019pos} considered bino decays taking place at various fixed target experiments in the ALPino model,\footnote{We extend their results by considering FASER, MATHUSLA, and NuCal detectors, and by also investigating the scattering LLP signatures described in \cref{sec:LLP_signatures}.} while ref. \cite{Bjorken:1988as}, discussing the results of the SLAC beam dump experiment E-137, considered its decay into gravitino and a photon.\footnote{It was an electron beam dump experiment, so the leading production channel relied on the t-channel selectron exchange. In light of LEP results on the masses of such states, the bounds on gravitino mass obtained in \cite{Bjorken:1988as} are not competitive with LEP limits \cite{DELPHI:2003dlq}, while we will show that FASER2 and SHiP may improve them for neutralinos with masses below $1\,\gev$.}
Other relevant works considering $\sim$sub-GeV neutralinos at the intensity frontier are \cite{Choudhury:1999tn,Dedes:2001zia,Dreiner:2002xg,Dreiner:2009er,Dreiner:2020qbi}.

Our study is motivated by recent developments of the far-forward LHC detectors \cite{Trojanowski:2023eru}, in particular the FASER experiment \cite{Feng:2017uoz,Feng:2017vli,FASER:2022hcn}, which began operation with the LHC Run 3.
In particular, a number of papers \cite{Dienes:2023uve,Kling:2022ehv,Dreiner:2022swd,Liu:2023nxi,Jodlowski:2020vhr,Jodlowski:2023yne,Jodlowski:2023sbi,Jodlowski:2023ohn} considered the prospects of multiple BSM scenarios using the signature of LLP decaying into a single photon and an invisible state.
Compared to the usual signature of LLP decay into a pair of charged leptons or photons, a single high-energy photon can be present in the detector, \eg, due to the muon or neutrino induced background.
However, dedicated studies performed by the FASER collaboration \cite{FASER:2022hcn} have shown that it will be sensitive to them with the same energy threshold on the visible energy deposited in the calorimeter by the decaying LLP; see also the extensive discussion in \cite{Jodlowski:2020vhr}.

In this work, we study such decays to investigate the prospects of sub-GeV bino-like neutralino at the LHC-based detectors: FASER2 \cite{FASER:2018ceo,FASER:2018bac,FASER:2021ljd}, FASER$\nu$2 \cite{FASER:2019dxq,FASER:2020gpr,Batell:2021blf,Anchordoqui:2021ghd}, and MATHUSLA \cite{Chou:2016lxi,Curtin:2018mvb}, as well as in the beam dump experiment CHARM \cite{CHARM:1985anb}, NuCal \cite{Blumlein:1990ay,Blumlein:2011mv}, and SHiP \cite{SHiP:2015vad,Alekhin:2015byh}.
We also consider additional LLP signatures - scattering with electrons and secondary bino production by upscattering of the LSP with nuclei taking place in front of decay vessel \cite{Jodlowski:2019ycu}. 
In particular, the latter process was shown to allow FASER2 to cover a significant portion of the shorter-lived LLP regime, characterized by LLP decay length of $d \sim 1\,\m$, for many BSM scenarios \cite{Jodlowski:2020vhr,Jodlowski:2023yne,Jodlowski:2023sbi,Jodlowski:2023ohn}.

The paper is organized as follows. 
In \cref{sec:models}, we introduce the SUSY models for low-energy PQ or SUSY breaking scales. 
We describe how such a regime naturally leads to long-lived neutralinos in both scenarios. 
In \cref{sec:neutralino_intensity_frontier}, we discuss the production mechanisms of light SUSY species - neutralino and LSP - in intensity frontier searches.
We also introduce signatures involving LLPs, and we describe details of their simulation.
In \cref{sec:results}, we discuss our results. The main one is the sensitivity reach of each detector considered for both neutralino decay scenarios.
In \cref{sec:conclusions}, we summarize our findings.

\begin{figure*}[tb]
  \centering
  \includegraphics[width=0.48\textwidth]{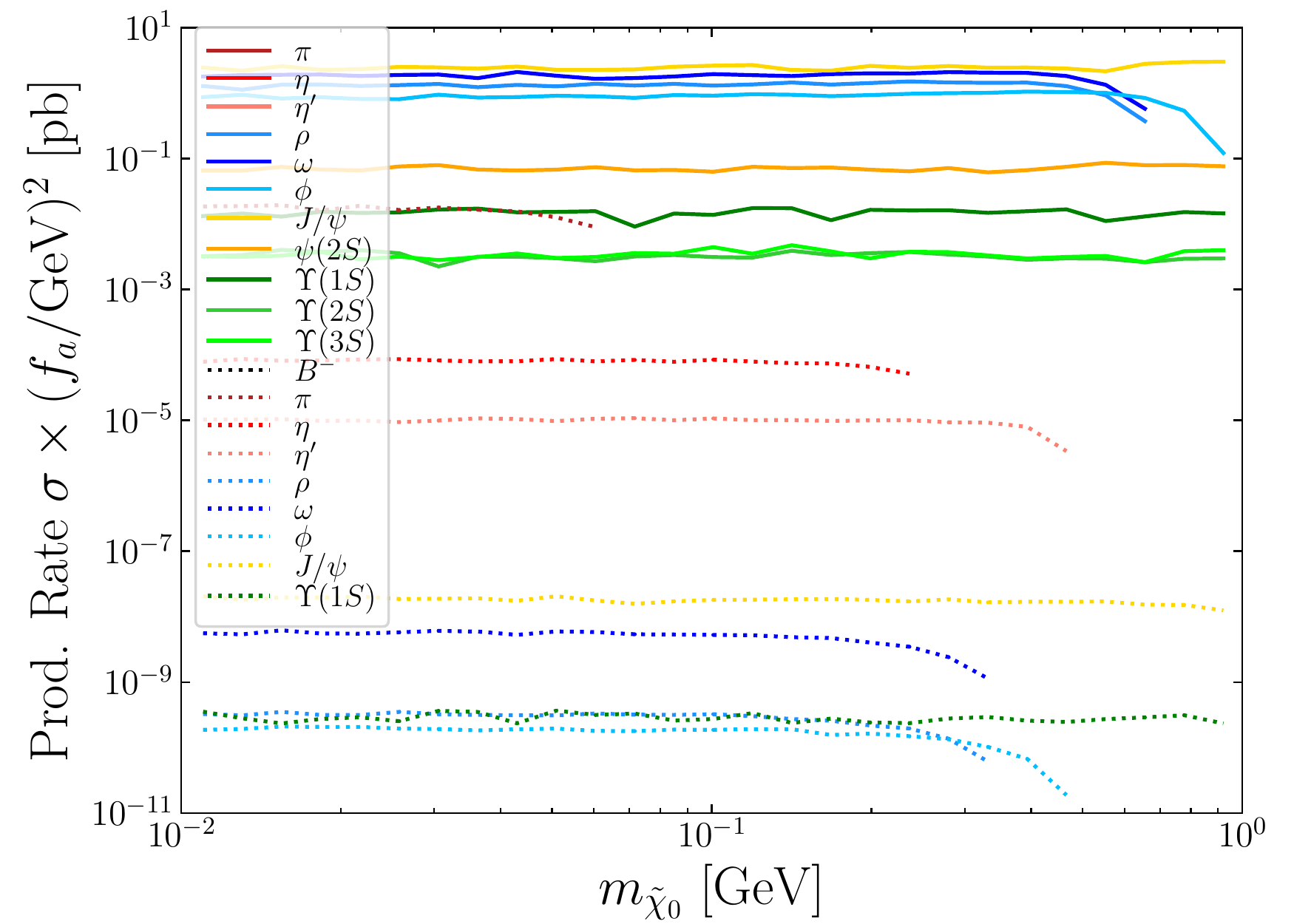}\hspace*{0.4cm}
  \includegraphics[width=0.48\textwidth]{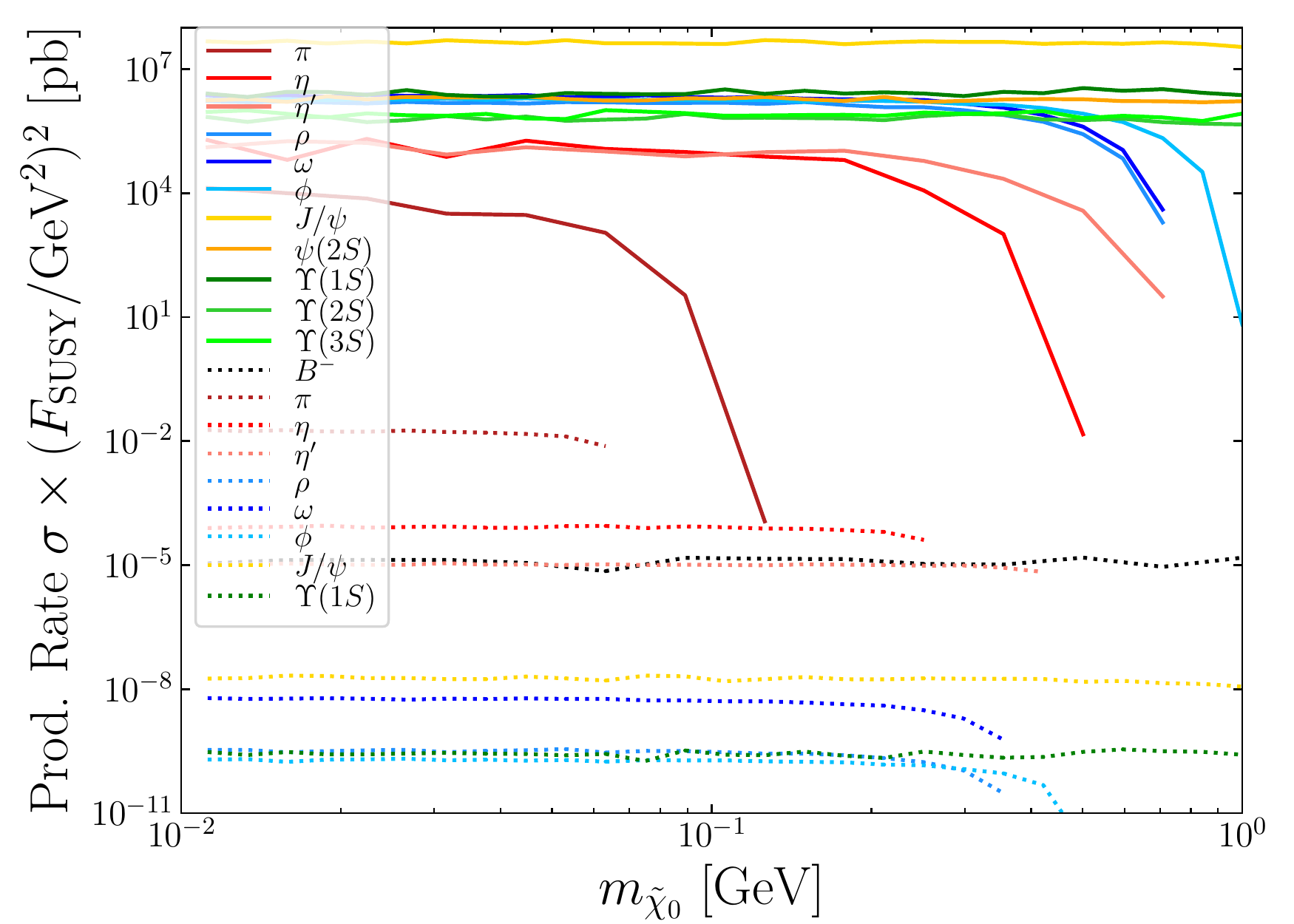}
  \caption{
    Modes of neutralino production by meson decays as a function of $m_{\tilde{\chi}_0}$ for ALPino (left) and gravitino (right).
    The decays into neutralino-LSP depending on $f_a$ or $F_{\mathrm{SUSY}}$ are denoted by solid lines, while decays into a pair of neutralinos, which are independent of such couplings, are indicated by dotted lines; color coding indicates contributions of each meson according to the legend.
    In fact, for ALPino, the $f_a$-independent decays dominate for the allowed values of $f_a$, while for gravitino, the decays depending on $F_{\mathrm{SUSY}}$ are the leading ones.
  }
  \label{fig:modes_production}
\end{figure*}

\section{Models\label{sec:models}}

Since we are interested in the $\mev$-$\gev$ mass range of a neutralino acting as the NLSP, we assume that it consists of pure bino, while the LSP is either ALPino or gravitino.\footnote{We note that a combined scenario with both ALPino and gravitino is possible and has interesting cosmological implications - see, \eg, \cite{Goto:1991gq,Chun:1993vz,Chun:1995hc}. In this scenario, NLSP decay into ALP and LSP is governed by SUSY breaking coupling, while ALP decays into SM states by PQ-scale suppressed couplings. Clearly, the LLP phenomenology crucially depends on the interplay between these two scales, which we leave for future study.}
In both of these models, the bino-LSP-photon coupling is the most relevant to our analysis, while other interactions - or specifics of mass spectrum of other SUSY states - play only a marginal role; see also the discussion in \cref{sec:prod_modes}.

Such a coupling is responsible not only for bino decays, $\tilde{\chi}_0 \to \mathrm{LSP} + \gamma$, but also for the leading production mechanism of $\tilde{\chi}_0$-$\mathrm{LSP}$ pairs - vector meson decays.
Moreover, it also governs the efficiency of the LSP-electron or LSP-nucleus upscattering processes that result in NLSP production. 
When such secondary production takes place in proximity to the decay vessel, it may allow the sensitivity reach to be extended to shorter-lived LLPs \cite{Jodlowski:2019ycu}.

The main similarity between our two SUSY scenarios is that the operator responsible for the $\tilde{\chi}_0$-$\mathrm{LSP}$-$\gamma$ coupling is mass dimension-5 operator, hence it is suppressed by the New Physics energy scale.
In fact, it is the PQ energy breaking scale $f_{a}$ for ALPino, and the SUSY breaking scale $\sqrt{F_{\mathrm{SUSY}}}$ for gravitino. 
Both of these parameters, along with the mass of the neutralino, are among the free parameters in our analysis.

On the other hand, while the ALPino mass is not necessarily strictly related to the SUSY or PQ breaking scales, and thus depends on the specifics of the SUSY scenario considered \cite{Chun:1992zk,Chun:1995hc,Choi:2013lwa}, the gravitino mass is fixed by the SUSY breaking scale, $m_{\tilde{G}} = F_{\mathrm{\mathrm{SUSY}}}/(\sqrt{3} \, m_{\mathrm{Pl.}})$, where $m_{\mathrm{Pl.}} = \sqrt{\hbar c/(8\pi G_N)} = 2.4 \times 10^{18}\,\gev$ is the reduced Planck mass, and $G_N$ is the Newton's gravitational constant.

\subsection{Neutralino-ALPino}
\label{sec:model_neutralino_ALPino}

The relevant part of the Lagrangian is \cite{Kim:1983ia,Kim:1984yn,Nieves:1986ed}
\be
  \!\!\mathcal{L} & \supset \frac{\alpha_{\mathrm{EM}} C_{a \gamma \gamma}}{16 \pi f_a} \overline{\tilde{a}} \gamma^5\left[\gamma^\mu, \gamma^\nu\right] \tilde{\chi}_0 F_{\mu \nu},
  \label{eq:L_chitilde_atilde}
\ee
where $\tilde{a}$ and $\tilde{\chi}_0$ denote the ALPino and neutralino fields, respectively, $F_{\mu \nu}$ is the electromagnetic (EM) field strength tensor, $\alpha_{\mathrm{EM}}$ is the fine structure constant, $C_{a \gamma \gamma}\sim O(1)$ is a mixing constant that depends on the ALP scenario \cite{Covi:1999ty,Covi:2001nw}, and $f_a$ denotes the PQ breaking scale.

The ALPino mass is in general a model-dependent quantity \cite{Chun:1992zk,Chun:1995hc,Choi:2013lwa}, so it essentially acts as a free parameter.
However, since the value of the ALPino mass will not significantly affect our discussion (as long as it is significantly smaller than the neutralino mass and does not cause large phase space suppression of the NLSP decay width), we follow \cite{Choi:2019pos} and set its value as follows: $m_{\tilde{a}}=10\,\mev$.

Ref. \cite{Choi:2019pos} considered the prospects of bino decays into ALPino and photon in various beam dump experiments, in which the NLSP production and decay points are separated by a large distance, typically $L \sim 100\,\m$.
This length scale largely determines the LLP decay length that can be probed is such detectors \cite{Bjorken:1988as,Bauer:2018onh}.
In the case of a sub-GeV bino, the following benchmark corresponds to a short-lived NLSP that can be covered in this way:
\be
  d_{\tilde{\chi}_0} \simeq &\, 100 \,\m \times \left(\frac{E}{1000\,\gev}\right) \left(\frac{0.1\,\gev}{m_{\tilde{\chi}_0}}\right)^4 \left(\frac{f_a}{30\, \gev}\right)^2,
  \label{eq:ctau_gtilde_atilde}
\ee
where $d_{\tilde{\chi}_0} = c \tau \beta \gamma$, $\tau = 1/\Gamma$ is the bino lifetime, $\gamma = E/m$ is the boost factor in the laboratory reference frame, and $\beta = \sqrt{1-1/\gamma^2}$.

The lifetime of a sub-GeV bino is determined by two-body decays given by \cref{eq:Gamma_axino_2body}, while three-body decays mediated by an off-shell photon typically contribute less than a percent, see \cref{eq:Gamma_axino_3body}.

\subsection{Neutralino-gravitino}
\label{sec:model_neutralino_ALPino}
The Lagrangian is \cite{Volkov:1972jx,Deser:1977uq,Wess:1992cp}\footnote{Feynman rules for this model are given in ref. \cite{Pradler:2006tpx}.}
\be
  \!\!\mathcal{L} & \supset -\frac{i}{8 m_{\mathrm{Pl.}}} \bar{\psi}_\mu [\gamma^\rho, \gamma^\sigma] \gamma^\mu \tilde{\chi}_0 F_{\rho \sigma},
  \label{eq:L_chitilde_gtilde}
\ee
where $\psi_\mu$ denotes the gravitino wavefunction, and the Lorentz index indicates the spin-$\frac 32$ character of the field.

Compared to the ALPino model, mass of gravitino is not a free parameter.
Instead, the SUSY breaking energy scale determines it by the super-Higgs mechanism \cite{Volkov:1973jd,Cremmer:1982en,Nilles:1982xx,Deser:1977uq,Cremmer:1978iv}. 
As a result, the gravitino mass is $m_{\tilde{G}} = F_{\mathrm{\mathrm{SUSY}}}/(\sqrt{3} \, m_{\mathrm{Pl.}})$.
Moreover, due to the SUSY Equivalence Theorem \cite{Casalbuoni:1988kv}, the gravitino wavefunction can be approximated at high energies as follows:\footnote{In our calculations, we instead take into the account all gravitino degrees of freedom through the gravitino polarization tensor given by \cref{eq:grav_tensor}.}
\be
  \psi_\mu \simeq i\sqrt{\frac 23} \frac{\partial_\mu \psi}{m_{\tilde{G}}},
\ee
where $\psi$ is the spin-$\frac12$ goldstino absorbed by the gravitino. 
As a result, even though gravitino interactions are suppressed by the Planck mass (due to its character as a SUSY partner of the graviton), cf. \cref{eq:L_chitilde_gtilde}, the massive gravitino compensates this suppression by the $\frac 1{m_{\tilde{G}}}$ factor.
Effectively, \textit{the bino-gravitino-photon coupling} is therefore proportional to the inverse of the SUSY breaking scale, $1/F_{\mathrm{SUSY}}$, instead of $1/m_{\mathrm{Pl.}}$.

For $\sim$sub-GeV neutralinos, the long-lived regime corresponds to low-energy SUSY breaking scales,
\be
  d_{\tilde{\chi}_0} \simeq &\, 100 \,\m \times \left(\frac{E}{1000\,\gev}\right) \left(\frac{0.1\,\gev}{m_{\tilde{\chi}_0}}\right)^5 \left(\frac{F_{\mathrm{\mathrm{SUSY}}}}{(60 \, \gev)^2}\right)^2,
  \label{eq:ctau_gtilde_Gtilde}
\ee
where $d_{\tilde{\chi}_0}$ is the bino decay length in the laboratory reference frame.
Its lifetime is determined by decays into gravitino and photon, while decays into gravitino and $e^+ e^-$ pair are suppressed, cf. \cref{eq:Gamma_grav_2body,eq:Gamma_grav_3body}; see the bottom panels of \cref{fig:results_gravitino}.

\section{Neutralino at intensity frontier searches\label{sec:neutralino_intensity_frontier}}

\subsection{Neutralino-LSP production modes \label{sec:prod_modes}}
Light neutralinos with masses below the mass of a proton can be efficiently produced in rare meson decays.
These mesons are generated in large numbers in both proton-proton (p-p) collisions at the LHC and proton-target collisions in beam dump experiments.

The branching ratio of vector and pseudoscalar meson decays into a pair of neutralinos taking place by $t$-channel squark exchange have been calculated in \cite{Borissov:2000eu,Dreiner:2009er}.
In fact, for the ALPino model, ref. \cite{Choi:2019pos} - see Eq. 9 and the discussion therein - used these results assuming the following mass spectrum: squark mass were set at $m_{\tilde{q}}=3\,\tev$, while the masses of the other SUSY particles were fixed at $10\,\tev$.\footnote{We adopt this SUSY mass spectrum in the following discussion. Reducing the value of any of these masses will not have a significant effect on our results, except for squark masses, which affect bino pair production, which would increase the detectors sensitivity to the ALPino model. However, it would typically not affect the gravitino scenario.}
As discussed in \cref{sec:results_plots_alpino}, this production channel allows for coverage of sizable part of the available parameter space.

On the other hand, in a number of BSM scenarios with a higher-dimensional LLP coupling to a photon, the leading LLP production channels are vector meson decays mediated by an off-shell photon, see, \eg, \cite{Merlo:2019anv,Chu:2020ysb,Jodlowski:2023sbi}.
Therefore, we also calculated the branching ratios of such decays, which are described by \cref{eq:brV} for the two SUSY scenarios considered.
We also considered the phase-space suppressed decays of pseudoscalar mesons - the corresponding differential branching ratios are given by \cref{eq:br2dq2dcostheta}.

As shown in \cref{fig:modes_production}, the leading decay channel among the photon-mediated decays is $J/\psi$ (gold solid line).
It is the heaviest meson among those produced in sufficiently large quantities - hence our result is consistent with the discussion in Sec. III in \cite{Chu:2020ysb}, which states that the vector meson branching ratio is approximately proportional to the mass squared of the meson.
The same relation holds for ALPino, while for gravitino, this branching ratio actually depends on the fourth power of the meson mass, cf. top and bottom lines of \cref{eq:brV}.

For the ALPino model, we found that indeed the dominant neutral meson decays are those mediated by the exchange of heavy squarks, which are thus independent of $f_a$, and result in the production of a neutralino pair. 
These contributions are denoted by dotted lines in \cref{fig:modes_production}, and for the allowed region of the parameter space, $f_a \gtrsim 200\,\gev$, they clearly overtake the photon-mediated decays (indicated by solid lines).

On the other hand, for the gravitino model, in the allowed SUSY breaking scale that is relevant for the intensity frontier searches, $200\,\gev \lesssim \sqrt{F_{\mathrm{SUSY}}} \lesssim 3\,\tev$, the photon-mediated decays dominate - see the right panel in \cref{fig:modes_production}.
Such a dependence can be explained by comparing the Lagrangians of the two models, given by \cref{eq:L_chitilde_atilde,eq:L_chitilde_gtilde}, respectively - the photon-coupling in the ALPino scenario has an additional factor of $\alpha_{\mathrm{EM}}/(2\pi) \simeq 10^{-3}$ with respect to the gravitino scenario.

We also checked that the pair production of neutralinos occuring in p-p collisions - either at the LHC or at beam dumps - does not improve the FASER or SHiP sensitivity towards the larger mass regime, $m_{\tilde{\chi}_0} \gg 1\,\gev$. 
The first detector has too small angular coverage, while the energy beam of the latter is too small to produce heavy neutralinos with sufficient abundance.
On the other hand, this production channel was found to allow MATHUSLA to cover long-lived very heavy neutralinos, $m_{\tilde{\chi}_0} \gtrsim 100\,\gev$ - see discussion in MATHUSLA physics case paper \cite{Curtin:2018mvb}. 
The difference between MATHUSLA and FASER is that the first of these LHC-based detectors has much larger decay vessel volume, is closer to the beamline, and is placed highly off-axis.
As a result, it covers a much larger part of the solid angle, and covers the LLPs produced with large transverse momentum $p_T$, while FASER is positioned in the far-forward direction.

Finally, we implemented the equations describing the production of bino-NLSP, \cref{eq:brV} and \cref{eq:br2dq2dcostheta}, and also the production of bino pairs, Eq. 9 from \cite{Choi:2019pos}, in a modified $\tt FORESEE$ package \cite{Kling:2021fwx}.
We use it to generate the spectrum of bino-NLSP, which is then used to simulate bino decays taking place in each detector considered and other signatures described in \cref{sec:LLP_signatures}.
We note that an opposite mass hierarchy between bino and ALPino or gravitino is also possible, and our simulation is adapted to such a case as well.

\subsection{Experiments and LLP signatures\label{sec:LLP_signatures}}
In this section, we briefly describe the intensity frontier detectors sensitive to single-photon LLP decays. 
We also introduce the main signatures of LLPs that we use to study our two SUSY scenarios.

The characteristics of both of these topics have been discussed at length in Sec. III in \cite{Jodlowski:2023sbi} for the dark axion portal, which is characterized by similar phenomenology.
In particular, in that scenario, the dark sector (DS) states are connected to the SM by a dimension-5 coupling to a photon.
Therefore, the following presentation is brief, while details can be found in the aforementioned work.

\paragraph{Experiments}
We consider a number of intensity frontier experiments in order to take advantage of their different features that may allow complementary coverage of the parameter space.
The specifics of each detector is given in Tab. 1 in \cite{Jodlowski:2023sbi}. We use these parameters in our simulation.

Among the beam dumps, these are: CHARM \cite{CHARM:1985anb}, NuCal \cite{Blumlein:1990ay,Blumlein:2011mv}, and SHiP \cite{SHiP:2015vad,Alekhin:2015byh}.
We also study detectors dedicated for LLP searches at the LHC, such as FASER2 \cite{FASER:2018ceo,FASER:2018bac,FASER:2021ljd}, FASER$\nu$ \cite{FASER:2019dxq,FASER:2020gpr}, and MATHUSLA \cite{Chou:2016lxi,Curtin:2018mvb}.

Moreover, as a result of the efforts of the LHC-based LLP community, another separate facility, called the Forward Physics Facility (FPF) \cite{MammenAbraham:2020hex,Anchordoqui:2021ghd,Feng:2022inv}, has been proposed.
It would contain not only a much larger version of the FASER2 detector, but it would also house additional ones. The detectors relevant to our analysis are FASER$\nu$2 \cite{Batell:2021blf,Anchordoqui:2021ghd} and FLArE \cite{Batell:2021blf}.

All of these detectors use beams of protons that hit a stationary, dense target in beam dumps or collide with other protons at the LHC.
Since the energy at the LHC is several orders of magnitude greater than that obtainable in beam dumps, FASER, MATHUSLA and other such detectors probe much more boosted LLPs than fixed target experiments.
On the other hand, the luminosity of the proton beam in dedicated beam dump experiments is significantly higher than at the LHC.
This usually results in their deeper reach toward smaller LLP coupling values.

\paragraph{Bino decays}
Our main LLP signature is bino decays occuring inside a decay vessel that is separated from the LLP production point by a distance $L\sim 100\,\m$. Such large separation, together with a dedicated infrastructure, allows to get rid of the SM background, perhaps except for neutrinos or muons, depending on the detector design.

Such decays take place with the following probability:
\be
  p(E) = e^{-L/d}(1 - e^{-\Delta/d}),
  \label{eq:p_prim}
\ee
where $\Delta$ is the detector length and $d=d(E)$ indicates the LLP decay length in the laboratory reference frame.

It is well known that the distance $L$ sets the length scale for the minimal decay length $d$ that can be probed in such a way \cite{Bjorken:1988as,Feng:2018pew,Bauer:2018onh}.
When $d \ll L$, $p(E)\simeq e^{-L/d}$, while in the opposite regime, $d \gg L$, $p(E)\simeq \Delta/d$ \cite{Essig:2013lka,Beacham:2019nyx}.

Consequently, only sufficiently long-lived species can be probed.
Since the detectors introduced in previous paragraph cover a wide range of the distance $L$, some variation in their sensitivity to LLP decays can be expected.

\paragraph{LSP-NLSP upcattering}
LLPs connected to the SM by a photon can also be efficiently produced by upscattering process of the lighter DS species into the heavier, unstable one - see Fig. 1 in \cite{Jodlowski:2019ycu} for an illustration, and \cite{Jodlowski:2019ycu,Jodlowski:2020vhr,Jodlowski:2023sbi,Jodlowski:2023ohn} for dedicated studies.

Such scattering is particularly efficient in the coherent regime, characterized by low-momentum exchange of the off-shell photon.
Then, it can take place in a coherent manner, scattering with the whole nucleus.

In fact, assuming a particular common from of the form factor - given in Eq. B2 in \cite{Jodlowski:2023yne} - which is responsible for the partial screening of the nucleus by the atomic electrons, one can obtain a closed-form of the upscattering cross section, $\mathrm{LSP} + N \to \mathrm{NLSP} + N$:
\be
  \sigma_{\tilde{a} N - \tilde{\chi}_0 N} \simeq & \frac{\alpha_{\mathrm{EM}}^3 \cos^2\theta_W Z^2}{16 \pi ^2 f_a^2} \times \\
  &\left(\log \left(\frac{d}{1/a^2 - t_{max}}\right) - 2\right), \\
  \sigma_{\tilde{G} N - \tilde{\chi}_0 N} \simeq & \frac{\alpha_{\mathrm{EM}} \cos^2\theta_W Z^2}{2 F_{\mathrm{SUSY}}^2} \times \\
  &\left( d + m_{\tilde{\chi}_0}^2\left(\log \left(\frac{d}{1/a^2 - t_{max}}\right) - 2\right) \right),
  \label{eq:Prim}
\ee
where $N$ is the nucleus, $m_e$ is the electron mass, $Z$ and $A$ are the atomic number and mass number of a nucleus, respectively, $a=111 Z^{-1/3}/m_e$, $d=0.164\,\gev^2 A^{-2/3}$, and $t_{\mathrm{max}} \simeq -(m_{\mathrm{NLSP}}^4+m_{\mathrm{LSP}}^4)/(4E_{\mathrm{LSP}}^2)$. 

We obtained these formulas by integrating the differential cross section following the method used in \cite{Dusaev:2020gxi} for photophilic ALP.
In the calculation of the squared amplitude, we included only the leading diagrams involving photon exchange. In particular, we neglected the diagrams with sleptons in the $t$-channel, which are negligible for slepton masses in the range of a few hundred GeV.

In our case, the production process takes place on tungsten (W) layers of the emulsion detector FASER$\nu2$ located upstream of FASER2, which is the main decay vessel.
Such upscattering depends on the coupling constant for $m_{\tilde{\chi}_0}=0.1\,\gev$ in the following way:
\be
  &\sigma^{\text{W}}_{\tilde{a} N \to \tilde{\chi}_0 N} \simeq \frac{1.5 \times 10^{-4}}{\mathrm{GeV}^2} \times \left(\frac{\mathrm{GeV}}{f_a}\right)^2, \\
  &\sigma^{\text{W}}_{\tilde{G} N \to \tilde{\chi}_0 N} \simeq \frac{3}{\mathrm{GeV}^2} \times \left(\frac{1\mathrm{GeV}^2}{F_{\mathrm{\mathrm{SUSY}}}}\right)^2.
  \label{eq:Prim_bench}
\ee
We note the suppression of the cross section for the ALPino is again caused by the $\alpha_{\mathrm{EM}}/(2\pi) \simeq 10^{-3}$ factor in the Lagrangian given by \cref{eq:L_chitilde_atilde} with respect to the gravitino scenario.
Moreover, even in the latter case, the upscattering cross section is not very efficient. In fact, comparing \cref{eq:Prim_bench} with the result obtained in \cite{Dusaev:2020gxi} for photophilic ALP, see Eq. 15 therein, it is smaller by a factor of $\sim 30$.

In the next section, we also describe the FASER2 sensitivity reach obtained by secondary neutralino production at FASER$\nu$2, followed by its decay inside either the neutrino detector, or inside the main decay vessel.
Then, instead of \cref{eq:p_prim}, we use Eq. 6 in \cite{Jodlowski:2023sbi} to obtain the probability of such decay - for details, see discussion therein.

In addition to the coherent scattering, incoherent scattering with protons or electrons is also possible. In fact, the latter signature is particularly useful for sub-GeV BSM species and it has been investigated in many papers, see, \eg, \cite{deNiverville:2011it,
deNiverville:2016rqh,Batell:2021blf,Kling:2022ykt}.

Proceeding analogously to the procedure used to derive \cref{eq:Prim}, we obtained the following closed forms of the electron upscattering cross sections:
\be
  \sigma_{\tilde{a} e^- \to \tilde{\chi}_0 e^-} \simeq & \frac{\alpha_{\mathrm{EM}}^3 \cos^2\theta_W}{16 \pi^2 f_a^2} \times \log\left(\frac{E_R^{\mathrm{max}}}{E_R^{\mathrm{min}}}\right), \\
  \sigma_{\tilde{G} e^- \to \tilde{\chi}_0 e^-} \simeq & \frac{\alpha_{\mathrm{EM}}\cos^2\theta_W}{2 F_{\mathrm{SUSY}}^2} \times \\
  & \left(  2 m_e (E_R^{\mathrm{max}}-E_R^{\mathrm{min}}) +  m_{\tilde{\chi}_0}^2 \log\left(\frac{E_R^{\mathrm{max}}}{E_R^{\mathrm{min}}}\right)  \right).
  \label{eq:Prim_e}
\ee

For this signature, instead of \cref{eq:p_prim}, we used Eq. 8 in \cite{Jodlowski:2023sbi}.
Since scattering signature is affected by large neutrino background, we adapt angular and energy cuts found in \cite{Batell:2021blf}, see also Tab. 1 in \cite{Jodlowski:2023sbi}.

\begin{figure}[tb]
  \centering
  \includegraphics[width=0.48\textwidth]{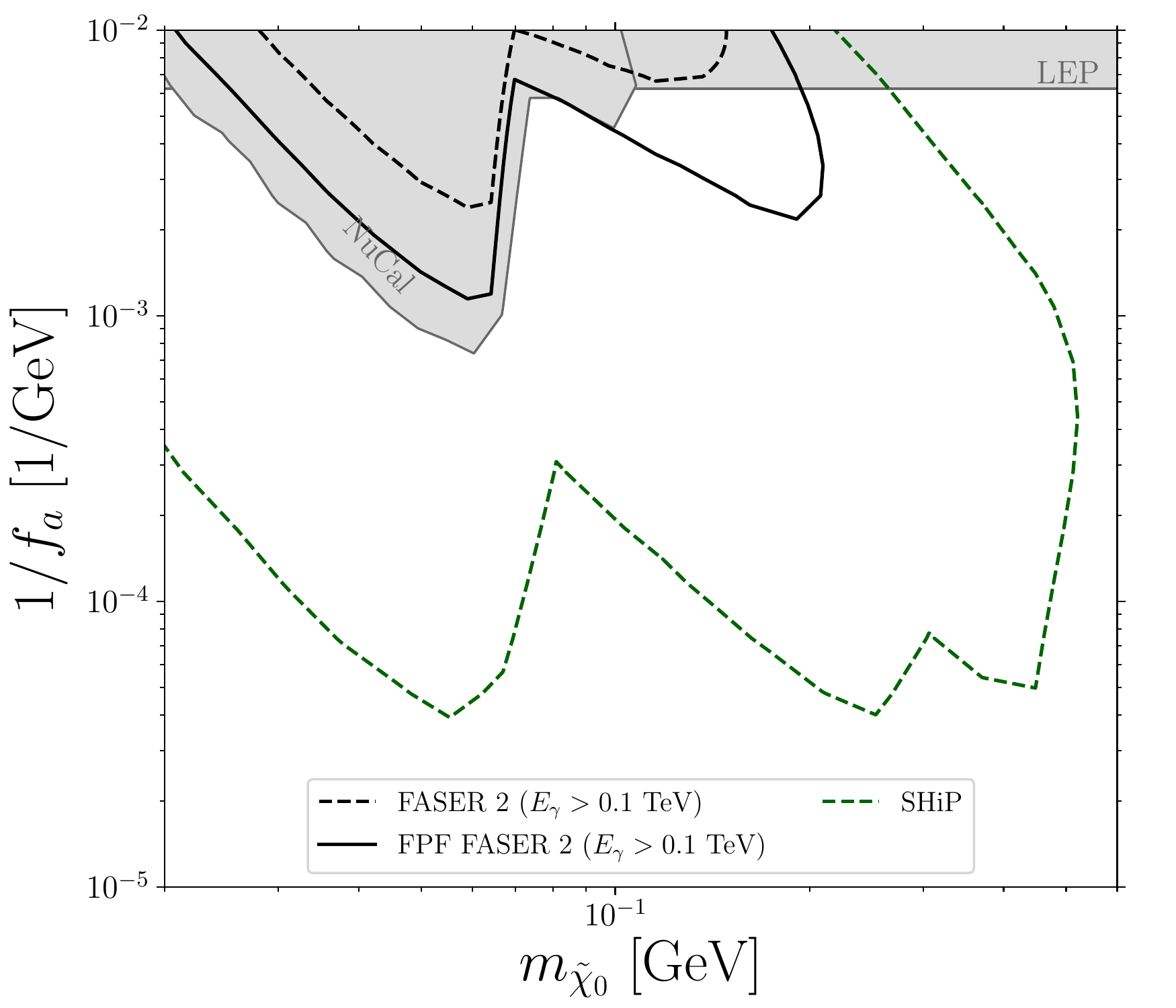}
  \caption{
    The sensitivity of FASER2 to neutralino decays into ALPino and photon for fixed $m_{\tilde{a}}=10\,\mev$. 
    The FPF version of the detector will exceed the current bounds set by NuCal and LEP due to its larger size compared to the baseline version of FASER2.
  }
  \label{fig:results_alpino}
\end{figure}

\begin{figure*}[tb]
  \centering
  \includegraphics[width=0.48\textwidth]{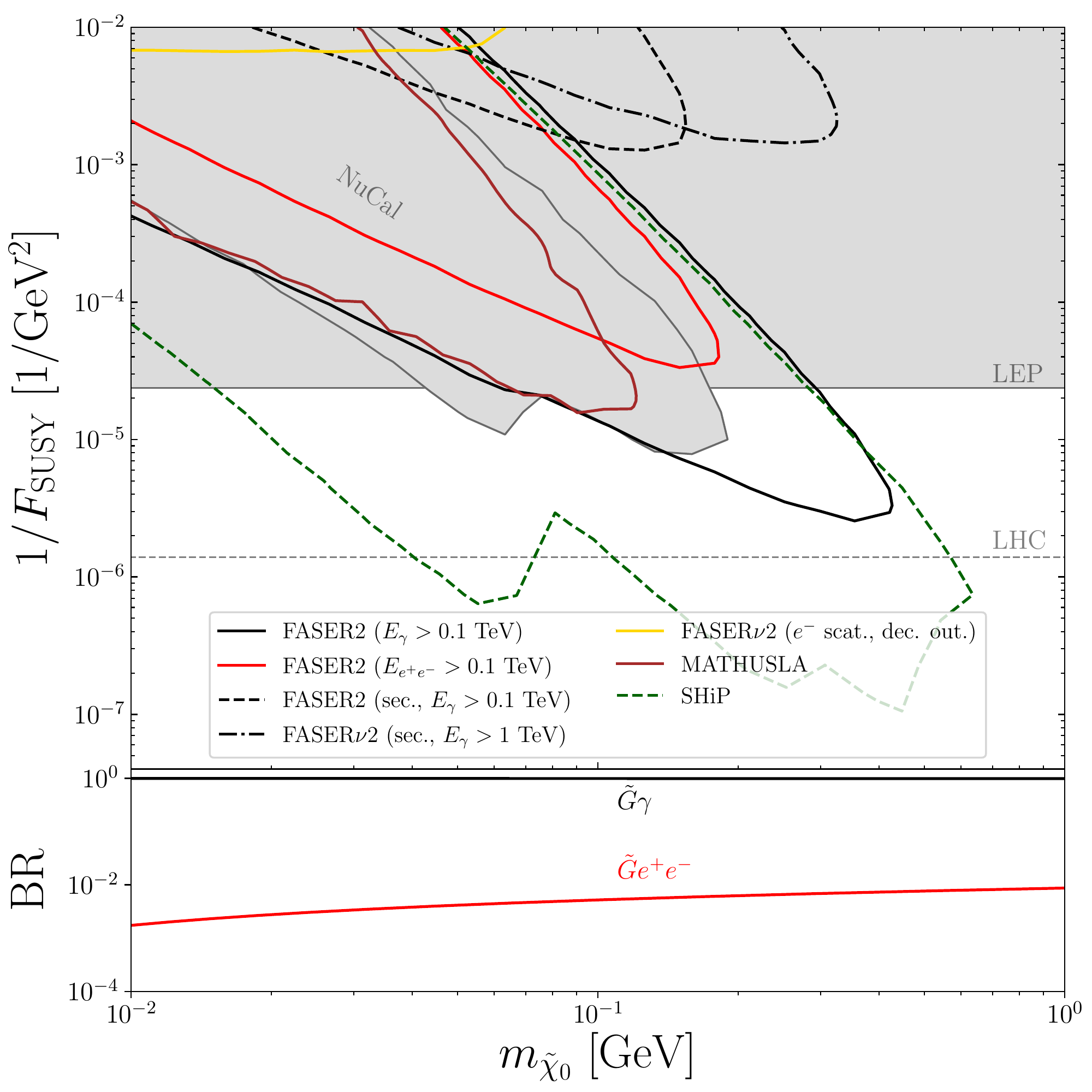}\hspace*{0.4cm}
  \includegraphics[width=0.48\textwidth]{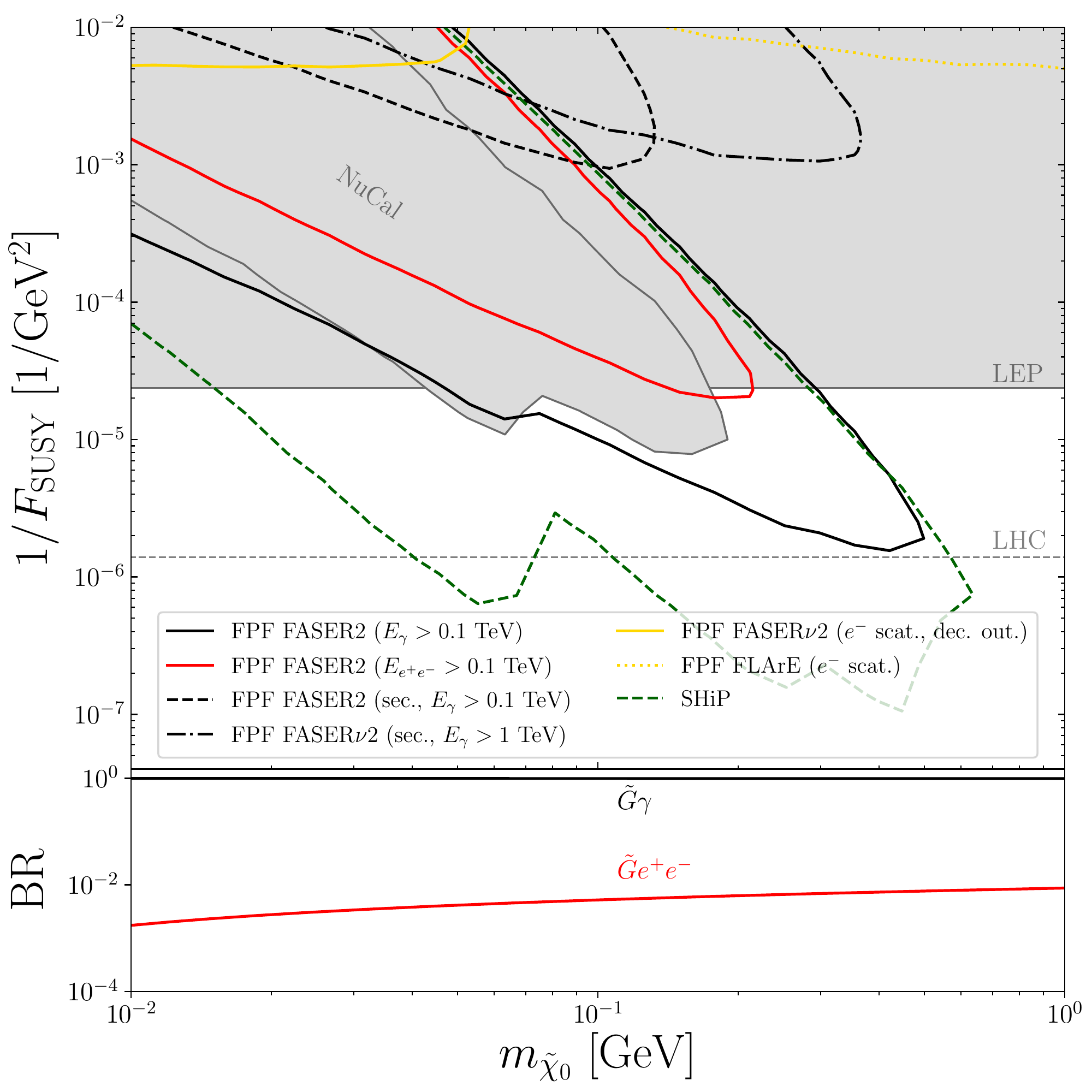}
  \caption{
    The sensitivity of FASER2, MATHUSLA, and SHiP to the neutralino-gravitino model.
    Leading two-body decays will allow FASER2 (solid black line) to partly extend the LEP bound, while FPF FASER2 will even reach the LHC (model-dependent) bound. 
    We also present results for three-body neutralino decays at MATHUSLA (brown) and FASER2 (red solid line), which cover high and low $p_T$ regimes of LLPs produced due to the p-p collisions at the LHC, respectively.
    Secondary neutralino production extends the sensitivity of FASER2 (black dashed and dot-dashed lines) into the short-lived, higher mass regime, while electron scattering at FASER$\nu$2 and FLArE (solid and dotted gold lines, respectively) covers the lower mass regime, which, however, are both already excluded by LEP.
  }
  \label{fig:results_gravitino}
\end{figure*}

\section{Results\label{sec:results}}

\subsection{Sensitivity reach for ALPino\label{sec:results_plots_alpino}}
In \cref{fig:results_alpino}, we present our results for the scenario when ALPino is the LSP.

For beam dump experiment, we find agreement with results of \cite{Choi:2019pos}. We consider additional detector of this type, NuCal, and we find that it actually improves over the NOMAD \cite{NOMAD:1997pcg} sensitivity shown in that work.

Since the leading channel of NLSP-LSP production is $f_a$-independent meson decay into a pair of binos, there is hardly any flux of ALPinos - see the discussion in \cref{sec:prod_modes}, in particular the left panel of \cref{fig:modes_production}.
As a result, neither the secondary production given by the top line of \cref{eq:Prim}, nor upscattering on electrons, given by the top line of \cref{eq:Prim_e}, are efficient.
Consequently, FASER2 will not have sensitivity to such signatures.

On the other hand, bino-pair production can be quite efficient.
While the baseline versions of FASER2 taking data during the High Luminosity era of the LHC will not improve over NuCal (but its sensitivity is greater than NOMAD), the FPF FASER2 will extend it in the $m_{\tilde{\chi}_0} \simeq 0.1\,\gev$ mass regime.
For LLP decays produced in the primary production, its main advantage over the baseline version is simply its larger size.

\subsection{Sensitivity reach for gravitino\label{sec:results_plots_Gravitino}}
On the other hand, when gravitino acts as the LSP, the dominant production modes produce equal fluxes of gravitinos and neutralinos, allowing the additional upscattering signatures described in \cref{sec:LLP_signatures}.
In fact, contrary to the ALPino scenario, both neutralino production and decays are controlled by the NLSP-LSP-photon coupling, which here depends on the SUSY breaking scale as $1/F_{\mathrm{SUSY}}$. 

This allows one to search not only for the displaced $\tilde{\chi}_0$ decays, but also for the electron scattering signature and for the decays of $\tilde{\chi}_0$ produced by upscattering occuring at the FASER$\nu$2 detector located before FASER2.

In \cref{fig:results_gravitino}, we present our main results for this model.
The areas shaded in gray are excluded by NuCal or LEP \cite{DELPHI:2003dlq,Mawatari:2014cja}. We also indicate a model-dependent bound from the LHC \cite{ATLAS:2012zim,deAquino:2012ru} by dashed gray line.

As mentioned earlier, we consider two versions of the FASER2 detector - the results for the baseline version are in the left panel, while the results for FPF FASER2 are in the right panel.
The sensitivity lines derived for the two-body bino decays are marked by black lines for FASER and green for SHiP, while those for three-body decays are indicated by red (FASER2) and brown (MATHUSLA) lines. 
The sensitivity lines correspond to the number of bino decays (number of LLP signatures in general case) given in Tab. 1 in \cite{Jodlowski:2023sbi} for each detector considered.

As is clearly seen, FASER2 will be able to significantly extend the LEP limit for $m_{\tilde{\chi}_0} \gtrsim 0.1\,\gev$ mass range, while searches for $e^+ e^-$ pairs produced in the three-body decays at MATHUSLA and FASER2 will be competitive with current LEP and NuCal bounds.
Moreover, the FPF version of FASER2 may even reach the strongest limit on light gravitinos coming from the LHC.

The upscattering signatures allow to cover the smaller lifetime regime, $d_{\tilde{\chi}_0}\sim 1\,\m$, which, however, is already excluded by LEP for both locations of the bino decays: FASER2 (black dashed line) and FASER$\nu$2 (black dot-dashed line).
Finally, the electron scattering signature covers the low mass region of the bino, which, however is also already excluded.

\section{Conclusions\label{sec:conclusions}}
Neutralino can act as a LLP decaying into a single photon and LSP in various SUSY scenarios.
Constraining them can be challenging in high-energy detectors designed for much heavier and short-lived BSM species. 
However, searches at the intensity frontier are well suited to such a regime.

We found that FASER2 and SHiP, which are particularly predisposed to such a difficult signature \cite{deNiverville:2019xsx,Jodlowski:2020vhr,Choi:2019pos,Dreiner:2022swd}, will be able to meaningfully extend the current constraints on the low-energy SUSY breaking scale in two scenarios involving sub-GeV neutralinos.
In the former, the LSP is composed of ALPino, while in the latter case, gravitino plays that role.

For the gravitino model, we also investigated additional LLP signatures: secondary neutralino production due to upscattering taking place at the FASER$\nu$2 detector in front of the main decay vessel, FASER2, and three-body decays depositing visible energy into a $e^+ e^-$ pair.

Finally, we considered the extended version of the FASER2 experiment, the proposed FPF. 
Due to its larger size, the FPF scenario of FASER2 may reach the strongest limit on light gravitinos coming from the LHC, while SHiP, due to its higher luminosity, may improve them further.

\acknowledgments

This work was supported by the Institute for Basic Science under the project code, IBS-R018-D1.

\appendix

\section{Neutralino decays}
\label{app:decays}

In these appendices, we give the relevant decay widths and cross sections, which we used in our simulation.

As in the rest of our paper, we assume the neutralino is composed of pure bino, which leads to an additional factor of $\cos\theta_W$ - where $\theta_W$ is the Weinberg angle - when reading the Feynman rules described by the Lagrangians given by \cref{eq:L_chitilde_atilde,eq:L_chitilde_gtilde}.

\subsection{ALPino}
The two-body decay width for bino decaying into an ALPino and a photon is \cite{Choi:2019pos}
\be
  \Gamma_{\tilde{\chi}_0 \to \tilde{a}\gamma} = \frac{\alpha_{\mathrm{EM}}^2 \cos^2\theta_W}{128 \pi^3} \frac{m_{\tilde{\chi}_0}^3}{f_a^2}\left(1-\frac{m_{\tilde{a}}^2}{m_{\tilde{\chi}_0}^2}\right)^3.
  \label{eq:Gamma_axino_2body}
\ee
Below, we give the integrated decay width for the leading three-body decay into an ALPino and an electron-positron pair in the limit of $m_{\tilde{\chi}_0} \gg m_{\tilde{G}}, m_{e^-}$,
\begin{dmath}[labelprefix={eq:}]
  \Gamma_{\tilde{\chi}_0 \to \tilde{a} e^+ e^-} \simeq \frac{\alpha_{\mathrm{EM}}^3 \cos^2\theta_W}{1152 \pi^4 f_a^2 m_{\tilde{\chi}_0}^3} \left( 18 m_{\tilde{\chi}_0}^4 m_{e^-}^2 - 4 m_{\tilde{\chi}_0}^6 - 32 m_{e^-}^6 + 3 m_{\tilde{\chi}_0}^6 \log \left(\frac{m^2_{\tilde{\chi}_0}}{4 m_{e^-}^2}\right) \right).
  \label{eq:Gamma_axino_3body}
\end{dmath}
The amplitude squared for the general mass scheme and the code evaluating such decay width, can be found in the auxiliary materials of the paper.

\subsection{Gravitino}
The two-body decay width for bino decaying into an gravitino and a photon is
\be
  \Gamma_{\tilde{\chi}_0 \to \tilde{G}\gamma} =& \frac{\cos^2\theta_W m^5_{\tilde{\chi}_0}}{16 \pi F_{\mathrm{SUSY}}^2} \left(1-\frac{m^2_{\tilde{G}}}{m^2_{\tilde{\chi}_0}}\right)^3 \left(1+\frac{m^2_{\tilde{G}}}{m^2_{\tilde{\chi}_0}}\right).
  \label{eq:Gamma_grav_2body}
\ee
We used the Feynman rules described in \cite{Pradler:2006tpx}, where in particular we used the full form of the gravitino polarization tensor.
It is defined as the sum of the gravitino field with momentum $p$ over its spin degrees of freedom, 
\be
  \Pi^{\pm}_{\mu\nu}(k) \equiv \sum_{s=\pm\frac 12, \pm\frac 32} \psi^{\pm, s}_\mu(k) \overline{\psi}^{\pm, s}_\nu(k).
  \label{eq:grav_tensor}
\ee
In the high-energy limit, where ${\Pi^{\pm}_{\mu\nu}(k) \simeq -\slashed{k}(g_{\mu\nu} -\frac{2p_\mu p_\nu}{3m^2_{\tilde{G}}})}$, we obtain the well-known result \cite{Ellis:2003dn,Giudice:1998bp,Diaz-Cruz:2016abv}, 
\be
  \Gamma_{\tilde{\chi}_0 \to \tilde{G}\gamma} = \frac{\cos^2\theta_W m^5_{\tilde{\chi}_0}}{16 \pi F_{\mathrm{SUSY}}^2} \left(1-\frac{m^2_{\tilde{G}}}{m^2_{\tilde{\chi}_0}}\right)^3 \left(1+3\frac{m^2_{\tilde{G}}}{m^2_{\tilde{\chi}_0}}\right).
\ee

Since the MATHUSLA detector may not be sensitive to single-photon decays, we also considered phase-space suppressed decays into a gravitino and an electron-positron pair.
In the limit of $m_{\tilde{\chi}_0} \gg m_{\tilde{G}}, m_{e^-}$, the following formula describes it:
\be
  \Gamma_{\tilde{\chi}_0 \to \tilde{G} e^+ e^-} \simeq & \frac{\alpha_{\mathrm{EM}} \cos^2\theta_W m_{\tilde{\chi}_0}^5}{576 \pi^2 F_{\mathrm{SUSY}}^2} \times \\
  & \left(24 \log \left(\frac{m_{\tilde{\chi}_0}}{m_{e^-}}\right) - 25 - 12 \log (4)\right),
  \label{eq:Gamma_grav_3body}
\ee
while the general formula for the amplitude squared can be found in the Mathematica notebook linked to the paper.

\section{Pseudoscalar and vector meson decays}
\label{app:prod}

\subsection{Vector meson decays}
The following are formulas for vector meson decays mediated by an off-shell photon that result in the production of LSP-NLSP pair, $V(p_0) \!\to\! \gamma^*(p_1+p_2) \!\to\! \mathrm{LSP}(p_1) + \mathrm{NLSP}(p_2)$:
\begin{widetext}
  \be
    \label{eq:brV}
    \frac{{\rm BR}_{V \rightarrow \tilde{a}\tilde{\chi}_0}}{{\rm BR}_{V \rightarrow e^+ e^-}} \!=\! \cos^2\theta_W & \frac{\alpha_{\text{EM}} \left(m_V^2+2 (m_{\tilde{a}}-m_{\tilde{\chi}_0})^2\right) (m_V^2-(m_{\tilde{a}}+m_{\tilde{\chi}_0})^2) \sqrt{\left(-m_V^2+m_{\tilde{a}}^2+m_{\tilde{\chi}_0}^2\right)^2-4 m_{\tilde{a}}^2 m_{\tilde{\chi}_0}^2}}{128 \pi ^3 f_a^2 \sqrt{m_V^2-4 m_e^2} \left(m_V^3+2 m_V m_e^2\right)}, \\
    \frac{{\rm BR}_{V \rightarrow \tilde{G} \tilde{\chi}_0 }}{{\rm BR}_{V \rightarrow e^+ e^-}} \!=\! \cos^2\theta_W & \frac{ (m_V^2 - (m_{\tilde{G}}+m_{\tilde{\chi}_0})^2) \sqrt{\left(-m_V^2+m_{\tilde{G}}^2+m_{\tilde{\chi}_0}^2\right)^2-4 m_{\tilde{G}}^2 m_{\tilde{\chi}_0}^2}}{8 \pi F_{\mathrm{SUSY}}^2 \alpha_{\text{EM}} \sqrt{m_V^2-4 m_e^2} \left(m_V^3+2M  m_e^2\right)} \times \\
    & \times \left(2 m_V^2 \left(m_{\tilde{G}}^2+m_{\tilde{G}} m_{\tilde{\chi}_0}-m_{\tilde{\chi}_0}^2\right)+m_V^4+(m_{\tilde{G}}-m_{\tilde{\chi}_0})^2 \left(3 m_{\tilde{G}}^2+m_{\tilde{\chi}_0}^2\right)\right),
  \ee
\end{widetext}
where ${\rm BR}_{V\rightarrow e^+ e^-}$ is the branching ratio corresponding to decays into $e^+ e^-$ \cite{Workman:2022ynf}, which we took from the PDG \cite{Workman:2022ynf}.

\subsection{Pseudoscalar meson decays}
The following formulas describe the differential branching ratios of the pseudoscalar meson decays into $\gamma(p_1)$ and $\mathrm{LSP}(p_2)$-$\mathrm{NLSP}(p_3)$ pair.
We use the form particularly useful for Monte Carlo simulations, where $q^2 =(p_2+p_3)^2$ is the momentum squared of the off-shell photon mediating the decay, and $\theta$ is the angle between the LSP momentum in the rest frame of the off-shell photon and the momentum of the off-shell photon in the meson rest frame.
\begin{widetext}
  \be
    \label{eq:br2dq2dcostheta}
    \frac{d{\rm BR}_{P \!\to\! \gamma \tilde{a} \tilde{\chi}_0}}{dq^2 d\cos\theta} = {\rm BR}_{P\rightarrow \gamma \gamma} \cos^2\theta_W &\!\times\!\! \Bigg[ \frac{\alpha_{\mathrm{EM}}^2 }{512 \pi ^4 f_a^2 m_P^6 q^6} \left(q^2 - m_P^2\right)^3 \sqrt{\left(m_{\tilde{\chi}_0}^2+m_{\tilde{a}}^2-q^2\right)^2-4 m_{\tilde{\chi}_0}^2 m_{\tilde{a}}^2} \\ 
    &\times \left((m_{\tilde{\chi}_0}+m_{\tilde{a}})^2-q^2\right) \left(\cos (2\theta) \left((m_{\tilde{\chi}_0}-m_{\tilde{a}})^2-q^2\right)+3 (m_{\tilde{\chi}_0}-m_{\tilde{a}})^2+q^2\right) \!\Bigg], \\    
    \frac{d{\rm BR}_{P \!\to\! \gamma \tilde{G} \tilde{\chi}_0}}{dq^2 d\cos\theta} = {\rm BR}_{P\rightarrow \gamma \gamma} \cos^2\theta_W &\!\times\!\! \left[ \frac{1}{64 \pi^2 F_{\mathrm{SUSY}}^2 m_P^6 q^6} (m_P^2-q^2)^3 (m_{\tilde{\chi}_0}^2-q^2)^4 (\cos (2\theta)+3) \!\right],
  \ee
\end{widetext}
where $m_P$ is the mass of pseudoscalar meson and ${\rm BR}_{P\rightarrow \gamma \gamma}$ is the branching ratio of the decay into two photons, which we took from the PDG \cite{Workman:2022ynf}.

\bibliography{main_photino}

\end{document}